# Proteomics Analysis of FLT3-ITD Mutation in Acute Myeloid Leukemia Using Deep Learning Neural Network

Christine A. Liang, MD, Lei Chen, MD, Amer Wahed, MD, Andy N.D. Nguyen, MD[*]

Department of Pathology and Laboratory Medicine, University of Texas Health Science Center at Houston, Texas, TX 77030

## Abstract

Deep Learning can significantly benefit cancer proteomics and genomics.  In this study, we attempt to determine a set of critical proteins that are associated with the FLT3-ITD mutation in newly-diagnosed acute myeloid leukemia patients.  A Deep Learning network consisting of autoencoders forming a hierarchical model from which high-level features are extracted without labeled training data. Dimensional reduction reduced the number of critical proteins from 231 to 20. Deep Learning found an excellent correlation between FLT3-ITD mutation with the levels of these 20 critical proteins (accuracy 97%, sensitivity 90%, specificity 100%). Our Deep Learning network could hone in on 20 proteins with the strongest association with FLT3-ITD.  The results of this study allow a novel approach to determine critical protein pathways in the FLT3-ITD mutation, and provide proof-of-concept for an accurate approach to model big data in cancer proteomics and genomics.

Key Words: AML; FLT3-ITD; Proteomics; Deep Learning; Neural Network

## INTRODUCTION

Acute myeloid leukemia (AML) is a neoplasm of the bone marrow which is caused by mutations in the myeloid stem cells leading to the formation of aberrant myeloblasts. The highly proliferative cancer cells impede the formation of normal blood cells, eventually causing death if patients are left untreated.  There are about 19,000 new cases and 10,000 deaths from this disease in 2016. As only a quarter of the patients diagnosed with AML survive beyond 5 years, there is an urgent need to find better treatments for this type of leukemia. AML includes many subtypes that share a common clinical presentation despite arising from diverse mutations and genetic events.  A variety of technologies targeting the gene, mRNA, microRNA and protein level have helped predicting the prognosis of AML patients. Interestingly, most AMLs only have only a few gene mutations, but prognosis of AML patients is quite varied. A possible explanation for this diversity is differences in protein signaling. The genetic aberrations and mutations of myeloid leukemic cells often cause a profound impact on the cellular protein networks. Proteomics include a vast collection of techniques allowing for analysis of proteins at the cellular level. Therefore, proteomics could be an ideal tool for predicting responses as well as for monitoring targeted therapy. Much work remains to determine a critical set of proteins involved in any particular mutation before pathogenesis can be elucidated.

*Corresponding author
Andy N.D. Nguyen, MD
Department of Pathology and Laboratory Medicine
University of Texas Health Science Center at Houston
6431 Fannin Street, MSB 2.292, Houston, TX, 77030
(713) 500-5337; fax (713) 500-0712
Nghia.D.Nguyen@uth.tmc.edu

Our study focuses on one of the most clinically significant mutations in AML, the FMS-Like Tyrosine Kinase 3 (FLT3) gene mutation. The FLT3 protein is a member of the class III receptor-tyrosine kinase (RTK) family and it is encoded by a gene located on chromosome 13q12 and. FLT3 shares a high degree of structural homology with the KIT, FMS and PDGFR receptors[1]. FLT3 plays a critical role in normal growth and differentiation of precursor cells in bone marrow. Upon binding of ligand, the FLT3 receptor dimerizes at the plasma membrane, leading to autophosphorylation and activation of several downstream effector signaling cascades. These cascades include the RAS/MEK, PI3K/AKT/mTOR, and STAT-5 pathways, all of which are important in cell cycle progression, inhibition of apoptosis, and activation of differentiation. Mutant FLT3 is expressed at higher levels, and demonstrates ligand-independence, causing constitutive autophosphorylation and activation of downstream signaling. The most common type of FLT3 mutations is internal tandem duplication (ITD) mutation. It activates signal transduction pathways in the juxta-membranous region, which are typically not activated by ligand-stimulated wild-type FLT3 [1,2]. ITD mutation occurs in about 23% of patients with de-novo AML. While AML patients with FLT3-ITD have potential to achieve initial complete remission as those with wild-type FLT3, they have a higher relapse rate and poorer response to salvage therapy. At the presence, inhibitors of FLT3 have not been shown to improve overall survival.

DREAM, which stands for Dialogue for Reverse Engineering Assessment and Methods, is a platform for collaborative community studies that focus on developing computational tools to solve biomedical problems[3]. The DREAM Challenges crowd-source non-profit studies which are supported by contributors from universities, computer technology companies like IBM Research, non-profit organizations like Sage Bionetworks, biotechnology and pharmaceutical companies. DREAM and Sage Bionetworks offer open data access to participants who wish to solve complex problems. Since the first DREAM Challenge in 2006, their participants have presented numerous findings in leading biomedical journals. DREAM challenges leverage the wisdom of the crowd to develop innovative computational models, and making these methods available to the public. All insight gained during a challenge is stored on their Synapse web site[4] to be shared with the research community. The DREAM Challenges offer a wonderful source of various types of cancer for scientific research. The DREAM 9 Challenge (AML Outcome Prediction Challenge), hosted by Rice University, Houston, Texas, provides a unique source of data on AML patients which we utilized in this study.

The correlation between protein expressions and mutations in cancer cells plays an important role in clinical applications[5]. The protein expression profiles from samples of cancer patients may be compared to those from normal samples, allowing for studying the disease pathology. Machine learning classification techniques have been used to classify tissue samples into mutated type versus normal type. However, due to the high dimensions of protein expression data (i.e. the high number of protein in each sample) and the availability of only a relatively small number of samples for a given mutation, the analysis presents significant challenges to how to process such data. The first challenge is to reduce the number of proteins in such a way that ensures sufficient information to perform accurate classification, but at the same time eliminates superfluous information (background noise). Several solutions have been made available to address the high dimension problem, most of which perform feature space reduction by constructing key features either manually or in supervised ways. This feature space reduction, however, leads to methods that are typically not scalable. The second challenge involves small sample sets (i.e. a small number of training examples) making the problem difficult to solve and increasing the risk of over-fitting. We propose, in this paper, the use of Deep Learning methods based on unsupervised feature extraction to address the two challenges described above.

Most successful Deep Learning methods involve artificial neural networks, a family of models inspired by biological neural networks (the central nervous system, particularly the brain). In such an artificial neural network, artificial nodes (known as "neurons") are connected together to form a network mimicking a biological neural network. Warren McCulloch and Walter Pitts created a computational model for neural networks based on an algorithm called threshold logic[6] in 1943. Neural networks had not shown superior performance compared to other machine learning methods until the introduction of Deep Learning in 2006. The core concept of Deep Learning involves learning the hierarchical structure of data by initially extracting simple low-level features, which are progressively used to build up more complex features, capturing the underlying features of the data. A simple example is demonstrated for a facial recognition task, in which each pixel of the image may be represented at the input layer. The input data are compressed in the hidden layer into features such as "large eye" or "small nose." In other words, the input data of the face can be described using learned features with less information than is given in the original image. Such compressed data can then be used to represent the input data at the output layer, allowing the facial image to be reconstructed entirely from the learned features. We use stacked autoencoders which form a deep network capable of achieving unsupervised learning, a type of machine-learning algorithm which draws inferences from the input data and does not use labeled training examples. In contrast to previous methods of conventional neural network where data must be strictly categorized to provide the appropriate label for supervised learning, the unlabeled data in Deep Learning can be used in unsupervised training phase. The resulting features from all training sets are then used as a basis for the construction of the classifier.

In this study, we attempt to use Deep Learning which incorporates unsupervised feature training to find correlation between the FLT3-ITD mutation and levels of a set of critical proteins. To the best of our knowledge, unsupervised feature learning methods have not been applied to protein expression analysis. Our study also involves the use of conventional neural network, which we will compare to the performance of the Deep Learning network against.

## MATERIALS AND METHODS
**Materials:**
The data in this study were obtained from the DREAM 9 Challenge that includes patients' demographics, cytogenetics, selected gene mutation status, and proteomic data for 191 patients diagnosed with AML[3]. The data had been de-identified to exclude all personal information prior to release on the DREAM 9 Challenge web site. Patients were all newly-diagnosed and had not been treated before blood sample drawing. Proteomics data include serum level of 231 proteins obtained by reverse phase protein array (RPPA) method.   The testing procedure has been described in details elsewhere[7] and is illustrated in Fig. 1. To exclude any factors that may confound the analysis in this study, we only included patients with normal cytogenetics, and FLT3-ITD (if present) is the sole mutation. With these restrictions, the number of cases in this study is reduced to 62 (normal cytogenetics, positive or negative for FLT3-ITD, no other mutations found).

Fig. 1 The Reverse Phase Protein Array (RPPA) Method to Measure Protein Levels[7]
Image Source[3]

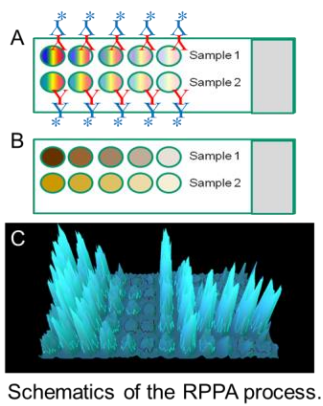

 Notes:
  - Cells were standardized at a concentration of 10,000 cells per microliter
  - Samples came from either blood or bone marrow. No statistical difference was
    determined for protein expression from blood vs. bone marrow

**Methods:**
Our main analysis method is a Deep Learning neural network with stacked (multi-layered) auto-encoder. Training will be mostly based on unsupervised feature learning which has been used successfully for image and audio recognition[8,9]. Our Deep Learning neural network was designed with the R language. R is a programming language for statistical computing and graphics supported by the R Foundation for Statistical Computing[10]. R was derived from the S language which was originally developed at Bell

Laboratories by John Chambers and colleagues. R's popularity has increased substantially in recent years with advances in machine learning[11]. The source code for the R software environment is written primarily in Java, C, FORTRAN, and also R itself. R is freely available under the GNU General Public License, and pre-compiled binary versions are provided for various operating systems including UNIX, Windows and MacOS. In this study, we use many Deep Learning functions obtained from various R packages which are available from the Comprehensive R Archive Network[12]. To compare the performance of our Deep Learning neural network to that of a conventional neural network, we will also include in this study a conventional neural network (EasyNN, Neural Planner Software, Cheshire, England),

Fig 2. A Conventional Neural Network with only Supervised Training Phase
Image Source[50]

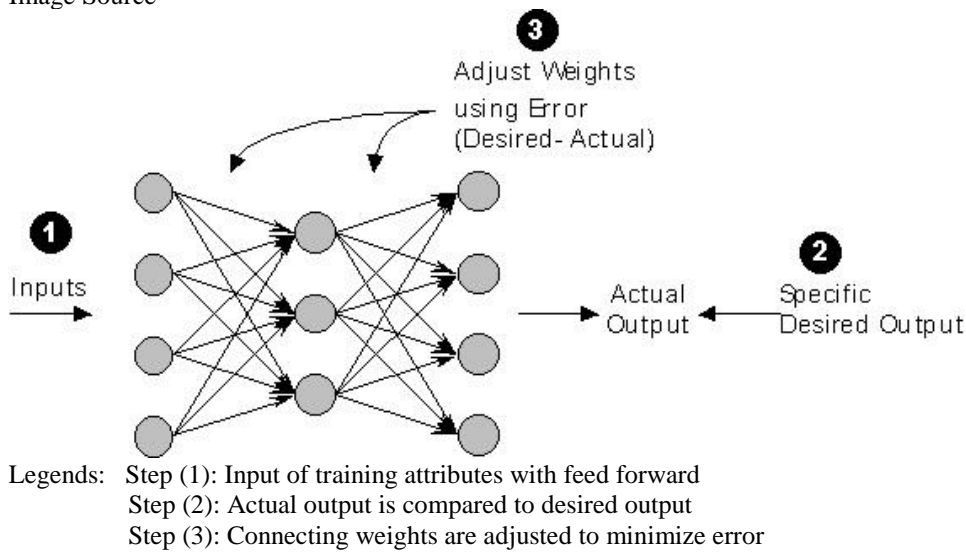

Legends:   Step (1): Input of training attributes with feed forward
           Step (2): Actual output is compared to desired output
           Step (3): Connecting weights are adjusted to minimize error

The conventional neural network, illustrated in Fig. 2, uses the well-established supervised training with back-propagation and sigmoid activation function[13]. The learning strategy for a conventional neural network starts with randomly initializing the weights of the network, followed by supervised backpropagation via gradient descent. This method has been shown to find suboptimal solutions for networks with multiple hidden layers[14]. It has been suggested that with randomly initialized weights, the gradient-based training of supervised neural networks may get stuck in local minima or plateaus[15] and that it is difficult to find a solution with more layers. The stacked autoencoder neural network, illustrated in Fig. 3, incorporates two training phases: pre-training with unsupervised learning method, and fine-tuning which is similar to the supervised back-propagation in conventional neural network[16,17]. During pre-training phase, the output from one layer is subsequently used as the input for the next output layer. The output from each layer in essence represents an approximation of the input data constructed from a limited number of features represented by the hidden units of the

Fig 3. A Deep Learning Neural Network (Stacked Autoencoder Network) with Unsupervised Training in Pre-training Phase and Supervised Training in the Fine-tuning phase
Image Source[51]

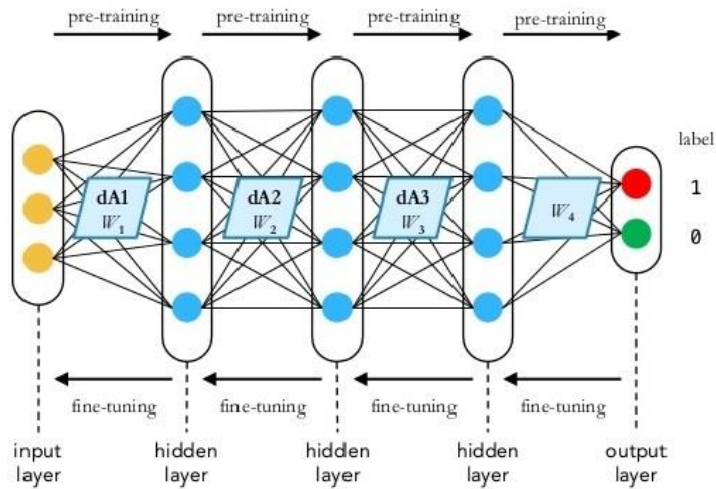

Legends:
dA1, dA2, dA3: autoencoders stacked in sequence
$W_1$, $W_2$, $W_3$, $W_4$: functions to extract key features from training attributes

network. The stacked autoencoder is constructed by multiple layers in the neural network (i.e. input layer, hidden layers, and output layer). The sigmoid function is used as activation function in hidden layers. In fine-tuning phase, the back-propagation method minimizes the error with an additional sparsity penalty[18]. The features learned in the pre-training phase are subsequently used with a set of labeled data for specific mutation status (positive or negative) to train a classifier. A classifier can be defined as a function that receives values of various features from training examples (protein levels as independent variables) and provides an output which predicts the category that each training example belongs to (mutation status as dependent variable)[19]. For the fine-tuning phase, we used linear function for the classifier.

We first performed training for both the conventional and the Deep Learning neural networks with the original training set including all 231 proteins and compared the two networks in terms of accuracy in predicting FLT3-ITD mutation status in the cross-validation sets. High dimensionality of protein expression data is likely to introduce background noise in addition to relevant proteins in the training set. We addressed this dimensionality problem in protein expression data by reducing the dimensionality of the feature space to the most relevant number of proteins based on the ranking of the proteins in training. The ranking of each protein is based on the sum of the absolute weights of the connections from the input node to all the nodes in the first hidden layer[20]. The performance of the two neural networks in terms of accuracy in predicting mutation status using this new scaled-down protein set was compared. A cross-validation method was used to obtain comprehensive validation results due to the small number of 62 samples. In this validation method, a small subset of data (10 out of 62) is excluded each time for training; the resultant trained network will be used to predict the mutation status for each case in the excluded subset. The process will be repeated until all 62 cases in the data set have been validated. The overall accuracy of each neural network is the mean of those for all the validated subsets.

**RESULTS**

The initial use of the full attribute set of 231 proteins yields 72% accuracy for the conventional network. The Deep Learning network performs better at 81% accuracy. Using the top 20 proteins ranked in this initial trial (Table 1), the conventional network achieves a better accuracy of 87%. The best accuracy is obtained by the Deep Learning network with 20 proteins at 97%. This remarkable accuracy corresponds to a sensitivity of 90% and a specificity of 100% for predicting positive FLT3-ITD status of an AML case using the level of 20 proteins. Using a smaller or larger number of proteins than 20 do not yield better accuracy (data not shown) indicating that 20 is the optimal number of proteins for this study. It appears that fewer than 20 proteins contain insufficient data for prediction. Conversely, more than 20 proteins would introduce background noise, compromising accuracy. Scaling down the number of proteins in training significantly reduces the number of data points for analysis from 14,322 (for 231 proteins and 62 cases) down to 1,240 (for 20 proteins and 62 cases). The accuracy in predicting FLT3-ITD status with different protein data sets by conventional neural networks vs. Deep Learning networks is summarized in Table 2. The compression of original features through training is illustrated in Fig. 4. The original features show a spread-out pattern whereas the extracted features are more compact, indicating a higher level of representation.

During the course of network training, we have tried various configurations for the two neural networks to achieve optimal accuracy and have the following important observations:
(a) The conventional neural net performs best with only one hidden layer, a fact well known with this type of neural network which relies strictly on supervised learning and multiple hidden layers present difficulty in training, often leading to no convergence in training (no learning achieved). For this reason, we use only one hidden layer for the conventional neural network in this study. Despite this limitation, 2 validation subsets (subsets 1 and 5) still show no convergence with 231 proteins.
(b) The Deep Learning network performs very well with 3 hidden layers consisting of 20, 15, 10 nodes, respectively for the 20-protein set. However, suboptimal results are obtained with the 231-protein set. For this reason, we use only 2 hidden layers which contain 10 and 5 nodes, respectively for 231 proteins to achieve better performance.

The use of machine learning algorithms frequently involves careful tuning of learning parameters and other model parameters. This tuning often requires experience, and sometimes brute-force search[21]. The parameters for the optimal configurations used in our neural networks, obtained through trial and error, are as follows:
- The conventional neural network[20]: Learning rate: 0.6, Momentum: 0.8
- The Deep Learning network[22]: Learning rate: 0.5, Momentum: 0.5

Table 1. The List of the 20 Top-Ranking Proteins Used in Training

| Column | Input Name | Importance |
|---|---|---|
| 98 | INPPL1 | 38.2049 |
| 46 | CLPP | 36.8005 |
| 165 | CDKN1B | 33.8749 |
| 13 | BAD#pS155 | 33.2770 |
| 215 | TP53 | 32.9059 |
| 54 | DIABLO | 29.3504 |
| 171 | PTPN11 | 29.1863 |
| 97 | INPP5D | 28.4851 |
| 103 | JMJD6 | 28.1224 |
| 182 | SIRT1 | 28.0780 |
| 221 | VHL | 28.0422 |
| 8 | ATF3 | 27.3000 |
| 66 | ERBB2 | 27.0661 |
| 211 | TAZ#pS89 | 26.7620 |
| 124 | MET#pY1230_1234_1235 | 25.3951 |
| 5 | ARC | 24.3036 |
| 213 | TGM2 | 23.8800 |
| 120 | MAPT | 23.6661 |
| 22 | BIRC5 | 23.6211 |
| 94 | HSPB1 | 23.4504 |

Legends:
-Column: position of the protein in the dataset
-Input Name: name of protein
-Importance: the sum of the absolute weights of the connections from the input node to all the nodes in the first hidden layer[20].

Table 2. Accuracy in Predicting Positive FLT3-ITD Status with Different Protein Data Sets by Conventional Neural Networks vs. Deep Learning Networks

| Neural Networks | 231 Protein Data Set | | 20 Protein Data Set | |
|---|---|---|---|---|
| Conventional | Validation Set No. | Accuracy | Validation Set No. | Accuracy |
| | 1 | NC* | 1 | 80% |
| | 2 | 80% | 2 | 80% |
| | 3 | 60% | 3 | 90% |
| | 4 | 60% | 4 | 90% |
| | 5 | NC* | 5 | 90% |
| | 6 | 90% | 6 | 90% |
| | 7 | 70% | 7 | 90% |
| | Mean= | 72% | Mean= | 87% ** |
| Deep Learning | 1 | 80% | 1 | 100% |
| | 2 | 90% | 2 | 100% |
| | 3 | 70% | 3 | 80% |
| | 4 | 80% | 4 | 100% |
| | 5 | 80% | 5 | 100% |
| | 6 | 90% | 6 | 100% |
| | 7 | 80% | 7 | 100% |
| | Mean= | 81% | Mean= | 97% *** |

Legends:
*NC: no convergence in learning
** corresponding to sensitivity of 75%, and specificity of 93%
*** corresponding to sensitivity of 90%, and specificity of 100%

Fig. 4 Graphic Display Representing the Original Features (Left) and the more compact Extracted Features (Right) Obtained Through Pre-Training in Deep Learning

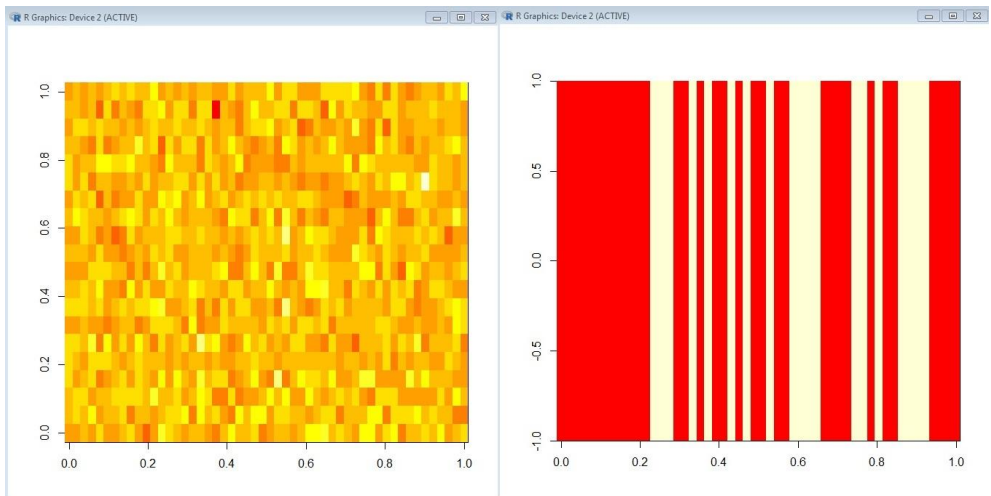

**DISCUSSION**

Deep Learning algorithms are new and innovative tools of research in machine learning to extract complex data representations at high levels of abstraction. In fact, Deep Learning has been cited as one of the 10 breakthrough technologies in 2013 by MIT Technology Review[23]. The most important contribution of Deep Learning algorithms is to develop a hierarchical architecture of data, where higher-level features are defined in terms of lower-level features. The hierarchical learning architecture of Deep Learning algorithms is motivated by the biological structure of the primary sensorial areas of the neocortex in the human brain, which automatically extracts abstract features from the underlying data[24-26]. Deep Learning algorithms rely on large amounts of unsupervised data, and typically learn data representations in a greedy layer-wise fashion[27,28]. Studies have shown that data representations obtained from stacking up nonlinear feature extractors (as in stacked autoencoders used in our study) often yield better machine classification results[29-31].

Deep Learning applications have produced outstanding results in different areas, including speech recognition[32-36], computer vision[27,28,37], and natural language processing[38-40]. A recent challenge hosted by the International Symposium on Biomedical Imaging (ISBI) in 2016 lead to a successful deep learning system for automated detection of metastatic cancer from whole slide images of sentinel lymph nodes[41]. Data-intensive technologies in proteomics and genomics as well as improved computational and data storage resources have contributed to Big Data science [42]. Technology-based companies such as Microsoft, Google, Yahoo, and Amazon have maintained databases that are measured in exabyte proportions or larger. Various private and public organizations have invested in Big Data Analytics to address their needs in business and research [43], making this an exciting area of data science research.

In the present study, we used Deep Learning for proteomics analysis in acute myeloid leukemia (AML). Specifically, we determined a set of 20 critical proteins that are associated with FLT3-ITD mutation out of 231 proteins available in newly-diagnosed AML patients. We implemented a Deep Learning network consisting of autoencoders that are stacked to form hierarchical deep models from which high-level features are compressed, organized, and extracted, without labeled training data. Dimensional reduction was initially performed to reduce the number of critical proteins from 231 to 20. We then showed how Deep Learning, which incorporates unsupervised feature training, can be used to find excellent correlation between positive FLT3-ITD mutation status with levels of these 20 proteins (an accuracy of 97%, sensitivity of 90%, and specificity of 100%). Our study also showed that the Deep Learning network outperforms the conventional neural network in this task (with lower accuracy of 86.7%, sensitivity of 75%, and specificity of 93%). Note that our objective is not to determine the set of critical proteins to detect FLT3-ITD mutation since existing testing technology with polymerase chain reaction (PCR) is much better for this purpose. Instead our goal is to determine what key proteins are involved in FLT3-ITD mutation.

**CONCLUSION**

The results of this study yield a critical dataset of 20 key proteins in FLT3-ITD mutation for further potential research to determine important protein pathways for this mutation in AML, to explore pathogenesis involving the mutation, to monitor chemotherapy response, and to design personalized treatment. To the best of our knowledge, Deep Learning with unsupervised feature learning methods has not been applied to protein expression analysis in AML. While the amount of data used here is relatively modest, this study provides a proof-of-concept for using Deep Learning neural network as a more accurate approach for modeling big data in cancer genomics and proteomics.

**Acknowledgement**: The data in this study were provided by Dr. Steven Kornblau from The University of Texas MD Anderson Cancer Center and were obtained through Synapse syn2455683 as part of the AML DREAM9 Challenge[3].